\documentclass{iopart}
\usepackage{graphicx}
\usepackage{iopams}
\usepackage{epstopdf}
\eqnobysec

\begin{document}
\title{Concurrence in Disordered Systems}
\author{Jenny Hide}
\address{The Abdus Salam International Centre for Theoretical Physics,
Strada Costiera 11, 34151 Trieste, Italy}
\ead{jhide@ictp.it}

\pacs{03.67.-a  05.70.-a  61.43.-j}
\date{\today}

\begin{abstract}

Quantum systems exist at finite temperatures and are likely to be disordered to some level.
Since applications of quantum information often rely on entanglement, we require
methods which allow entanglement measures to be calculated in the presence of
disorder at non-zero temperatures. We demonstrate how the disorder averaged concurrence
can be calculated using thermal many-body perturbation theory. Our technique can also be
applied to other entanglement measures. To illustrate, we find the disorder averaged
concurrence of an $XX$ spin chain. We find that concurrence can be increased by
disorder in some parameter regimes.

\end{abstract}

\maketitle

\section{Introduction}

Disorder is an unavoidable feature of many-body systems
\cite{CondMatFT,QFTStat,RandomSpinGBook}.
Since the properties of large disordered systems are difficult to study,
tools have been developed to tackle them such as averaging
over the disorder using sampling or perturbation theory,
or using renormalisation group techniques. For example,
strong disorder in $XXZ$ spin chains can be investigated using
renormalisation groups \cite{Dasgupta,Fisher}. Entanglement
is another important feature of many-body systems \cite{manybody}, one that
has been shown to be a useful resource in many quantum information and
computation schemes. It is therefore important to consider how
entanglement in real, finite temperature systems is affected by disorder.

In the context of entanglement, average disorder in spin chains has been studied
previously \cite{refael,lafl,santa,Chiara,random1,binosi,refael2,Hoyos}.
This research concentrated on using sampling, or on renormalisation
groups at zero temperature. In \cite{JHideDisW,JHideDMDis}, a
perturbative technique to calculate a disorder averaged (finite temperature)
entanglement witness was introduced. In this paper,
the perturbative method is again used, and it is demonstrated how
entanglement \textit{measures}, rather than a witness, can be calculated. While both measures and
witnesses are useful, considering entanglement measures allows any changes
in the \textit{amount} of entanglement by the disorder to be found.
The disorder averaged concurrence along with other entanglement measures for weakly
random systems are considered.

A physical realisation of a spin chain is highly unlikely to be free from random variations
of couplings or fields. Due to difficulties in cooling a system to its ground state, it is
also likely to be in a thermal, mixed state. Spin chains have been shown to be good
candidates for quantum wires, and thus have been studied extensively \cite{Bose, Bose2}.
The effects of a finite temperature \cite{Bayat,Bayat2,Wang} have been considered, as have
the effects of disorder at zero temperature \cite{Burgarth,Burrell,Burrell2,Allcock,Petrosyan,Wang},
however the combined effects of disorder and finite temperature have not. This paper proposes
how entanglement measures can be calculated for a random thermal spin chain.

Two distinct averages can be taken over the disorder: quenched or annealed \cite{RandomSpinGBook}.
Each is useful in differing situations, though the quenched average
is often considered the appropriate one \cite{experi}.
The quenched average corresponds to the disorder effectively being time independent;
the disorder is not in equilibrium with the system, and while the
system evolves, the disorder remains frozen. Calculation-wise, this requires
the thermal average to be taken before the average over the disorder, and thus
we must average over the logarithm of the partition function. In contrast,
the annealed average requires the average to be taken over the partition
function itself. In this case, the disorder changes quickly, and is in equilibrium
with the system as it evolves. Thus we take the thermal average and disorder
average at the same time. Of course, the quenched and annealed averages
are the two extremes; disorder with aspects of both can exist.

Many-body perturbation theory allows us to calculate disorder averaged
correlation functions which are crucial for quantifying entanglement.
As a consequence of the Jordan-Wigner transformation which transforms
qubits into spinless fermions, we can use fermionic perturbation theory to
study such spin systems \cite{PathIntSpinJW}. In this
paper, we calculate thermal disorder averaged correlation functions and
use them to find the disorder averaged concurrence.

Since the quenched or annealed average over disorder can be characterised by taking,
respectively, an average over the logarithm of the partition function or over
the partition function itself, we use the partition function as a starting
point to calculate the disorder averaged concurrence. Casting the partition function
into functional (or path) integral form, introducing a generating functional term,
and replicating it, we find we can take the average over the disorder, and later
calculate a perturbative expansion of the disorder averaged correlation functions
for both types of average \cite{CondMatFT,QFTStat}.
Thus rather than calculating a direct average over the concurrence, we construct
it from disorder averaged correlation functions.

In particular, we consider an $XX$ spin chain with a random term in the
thermodynamic limit. In addition to the disordered concurrence, we also
discuss the results for the disordered single and two site entanglement entropy
since by calculating the relevant disordered correlation functions for the
concurrence, we already have all the necessary ingredients for the calculation.

\section{Entanglement Measures}

A number of entanglement measures exist, and each has advantages
and disadvantages compared to the others. In this section, we
briefly discuss two different entanglement measures to demonstrate
how perturbation theory could be used to consider disorder in each of
them.

The concurrence, \cite{HillWoott,Wootters} quantifies entanglement
between two mixed qubits,

\begin{equation}
\mathcal{C} (\rho) = \rm{max} \{ 0, \lambda_1- \lambda_2-\lambda_3-\lambda_4\},
\end{equation}
where $\lambda_1 \geq \lambda_2 \geq
\lambda_3 \geq \lambda_4$. The $\lambda_i$'s are the square roots of the
eigenvalues of the matrix $\rho (\sigma^y
\otimes \sigma^y)\rho^* (\sigma^y \otimes \sigma^y)$, where $\sigma^y$ is a
Pauli spin matrix and $\rho^*$ is the complex conjugate of $\rho$. Concurrence
has an advantage over other entanglement measures since is is relatively easy
to calculate, and can be used for mixed states. Thus it can be used
to study thermal entanglement. However, concurrence is limited in that it can only be
used between pairs of qubits.

The entropy of entanglement \cite{EntanglementEntropy} measures the amount
of entanglement in a pure state, $\rho_{AB}$. It is defined as the von
Neumann entropy of a reduced density matrix,
$S\left( \rho_{A} \right)= -\mathrm{tr} \left( \rho_A \log \rho_A \right)$
where $\rho_A = \mathrm{tr}_B(\rho_{AB})$. The entanglement entropy measures
how mixed the subsystems, $A$ and $B$, of a bipartite system are. Although this
is only a measure of entanglement for pure states since for mixed states
it measures both quantum and classical correlations, it has the advantage that
it can be used to calculate entanglement between any two parts of a system.

Since both of these entanglement measures use the density matrix $\rho$ which
can be written in terms of correlation functions, we can use many-body perturbation
theory to find a perturbative expansion of each. For the concurrence, we only
need two qubits, $l$ and $m$, of a system:

\begin{equation}
\rho_{l,m} = \sum_{\alpha,\beta
= 1,x,y,z} \sigma_l^\alpha \sigma_m^\beta \langle \sigma_l^\alpha
\sigma_m^\beta \rangle,
\end{equation}
where $\langle \cdots \rangle = \mathrm{tr} (\rho \cdots)$ is valid
both for a thermal and a pure $\rho$. For two-site entanglement entropy,
we need $m=l+1$. The density matrix is even simpler if we wish
to consider single-site entanglement entropy, then $\rho_A = \sum_{\alpha
= 1,x,y,z} \sigma_l^\alpha \langle \sigma_l^\alpha \rangle$. Thus there are
fewer correlation function averages to calculate.

Other entanglement measures such as the relative entropy of entanglement,
$S(\rho || \sigma_{css}) = -\tr (\rho \log \rho) - \tr (\rho \log  \sigma_{css})$
where $ \sigma_{css}$ is the closest separable state to $\rho$, would be more
difficult to calculate. In this case, we would need to find $\sigma_{css}$
once the average had been taken.

In the remainder of this paper, we concentrate on calculating the disorder
averaged correlation functions necessary for finding the concurrence.

\section{The Model}

We consider an example, the XX spin chain in the thermodynamic limit, $N \rightarrow
\infty$, which means we can safely ignore boundary effects. This has Hamiltonian

\begin{equation}
H_0 = -\frac{J}{2} \sum_{l=1}^{N-1} \left( \sigma_l^x \sigma_{l+1}^x +  \sigma_l^y
\sigma_{l+1}^y \right) -B \sum_{l=1}^N \sigma_l^z,
\end{equation}
where J is the coupling strength between neighbouring spins and
B is an external magnetic field. $H_0$ is the
unperturbed part of the total Hamiltonian, $H=H_0+H_1(\mu_l)$ where $H_1(\mu_l)$ is the
perturbation, and $\mu_l$ is a random variable. We could use any distribution for the
random term, but we consider the case when $\mu_l=j_l$
is taken from a Gaussian distribution centred at zero with variance $\Delta$ and

\begin{equation}
H_1 (j_l) = -\frac{1}{2} \sum_{l=1}^{N-1} j_l \left( \sigma_l^x \sigma_{l+1}^x +
\sigma_l^y \sigma_{l+1}^y \right).
\end{equation}
We apply a Jordan-Wigner transformation, $a_l = \prod_{m=1}^{l-1} \sigma_m^z
\otimes (\sigma_l^x + i \sigma_l^y)/2$ to the total Hamiltonian to get

\begin{equation}
H = - \sum_{l=1}^{N-1} \left( J+j_l \right) \left( a^\dagger_l a_{l+1} + a^\dagger_{l+1} a_l \right) -
B \sum_{l=1}^N \left( \textbf{1}-2a^\dagger_l a_l \right).
\label{Eq:HSecondQ}
\end{equation}
A Fourier transform,
$a_l = \int_{-\pi}^\pi \frac{dq}{2\pi} e^{i q l} d(q)$ diagonalises the unperturbed
Hamiltonian, leaving $H_0 = \int_{-\pi}^\pi \frac{dq}{2\pi} \varepsilon (q) d^\dagger (q) d(q)$ where
$\varepsilon (q)= 2(B-J\cos q)$. Although we applied the Jordan-Wigner transformation to
$H_1$, the Fourier transform is not useful.
In order to treat $H_1$, we turn to fermionic many-body perturbation
theory using equation \ref{Eq:HSecondQ}.

The diagonal form of $H_0$ allows us to calculate useful quantities such as the partition
function, $Z_0= \tr [\exp(-\beta H_0)]$,

\begin{equation}
\ln Z_0 = N \int_{-\pi}^{\pi} \frac{dq}{2\pi} \ln \left\{ 2\cosh \left[ \frac{\beta \varepsilon (q)}{2}
\right] \right\}.
\label{Eq:PartFn0}
\end{equation}
where $\beta^{-1}$ is the temperature and we have set $k_B=1$.

\section{Concurrence}
\label{Sec:Conc}

As discussed previously, to take disorder into account when calculating concurrence,
we must find the disorder average of each correlation function. For the XX spin
chain many of the correlation functions are zero.
The concurrence between sites $l$ and $l+R$ in terms of correlation functions is
$\mathcal{C} = \frac{1}{2}\rm{max} \left\{0,| \langle \sigma_l^x \sigma_{l+R}^x+ \sigma_l^y
\sigma_{l+R}^y \rangle | - \sqrt{(1+ \langle \sigma_l^z \sigma_{l+R}^z \rangle )^2 -
4\langle \sigma_l^z \rangle^2} \right\}$. Thus for the disorder averaged concurrence,
we need to calculate $\overline{\langle \sigma_l^x
\sigma_{l+R}^x+ \sigma_l^y \sigma_{l+R}^y \rangle}$, $\overline{\langle \sigma_l^z
\sigma_{l+R}^z \rangle}$ and $\overline{\langle \sigma_l^z \rangle}$.
The bar indicates the average over the disorder while the brackets, $\langle \cdots \rangle
= \tr (\rho \cdots)$ denotes the thermal average. Each of these expectation values can
be calculated as a perturbative expansion in $\Delta$, the variance of the random
distribution, using a functional integral technique.
In this paper, we consider nearest and next nearest neighbour concurrence, where $R=1$
and $2$ respectively, however, the method we discuss applies to any value of $R$.

The expectation values for the unperturbed $XX$ chain were found in \cite{BarouchMc}.
We now outline how they are calculated for the unperturbed system before
discussing how to find their perturbative expansion. Using the Jordan-Wigner transformation,
the necessary correlation functions are $\langle \sigma_l^x \sigma_{l+R}^x +  \sigma_l^y
\sigma_{l+R}^y \rangle = 2\langle a_l^\dagger \prod_{m=l+1}^{R-1} (\textbf{1}-2 a_m^\dagger  a_m)
a_{l+R} + a_{l+R}^\dagger  \prod_{m=l+1}^{R-1} (\textbf{1}-2 a_m^\dagger  a_m) a_l\rangle$,
$\langle \sigma_l^z \sigma_{l+R}^z \rangle = \langle (\textbf{1}-2 a_l^\dagger  a_l)
(\textbf{1}-2 a_{l+R}^\dagger  a_{l+R})\rangle $ and
$\langle \sigma_l^z \rangle = \langle \textbf{1}-2 a_l^\dagger  a_l\rangle$.

Wick's theorem allows us to express n-point fermionic (and bosonic)
correlation functions in terms of two-point correlation functions when $H_0$ is a
non-interacting system. Thus to calculate each of the above, we need only one quantity,
$G_R = -\langle a_l^\dagger a_{l+R} - a_l a_{l+R}^\dagger \rangle $, remembering the
commutation relation, $\{ a_l^\dagger , a_m \} = \delta_{l,m}$.

Applying the Fourier transform we used to diagonalise the unperturbed
Hamiltonian, we get

\begin{equation}
G_R = \int_{-\pi}^\pi \frac{dq}{2\pi} \cos(qR) \langle 1-2 d^\dagger (q) d(q) \rangle.
\label{Eq:GR1}
\end{equation}
When $R=0$, this is equal to the magnetisation of a single site,
$M_0/N = \sum_l \langle \sigma_l^z \rangle/N$. Magnetisation is defined as $M_0=
\frac{1}{\beta} \frac{\partial}{\partial B} \ln Z_0$ and

\begin{equation}
\frac{M_0}{N} = \int_{-\pi}^\pi \frac{dq}{2\pi} \tanh \left[ \beta
\zeta (q) \right].
\end{equation}
Thus with the extra $\cos (qR)$ term in equation \ref{Eq:GR1}, we have

\begin{equation}
G_R = \int_{-\pi}^\pi \frac{dq}{2\pi} \cos(qR) \tanh \left[ \beta
\zeta (q) \right].
\end{equation}
Once we consider disorder, again due to Wick's theorem (since our $H_0$ is non-
interacting), we must find an
equivalent $\overline{G}_R$. Since all the correlation functions can be written
in terms of $\overline{G}_R$, it is sufficient to calculate
$\overline{\langle d^\dagger (q) d(q) \rangle}$.

\section{Functional Integrals}

We introduce the functional partition function (or partition
functional for convenience) which we will use to calculate the
expectation values for the disorder averaged concurrence:

\begin{equation}
Z\left[ \overline{\eta},\eta \right] = \tr \left[ \exp \left(-\int_0^{\beta} d\tau
\left\{ H-\sum_l \left[ \overline{\eta}_l(\tau) a_l + a_l^{\dagger} \eta_l
(\tau) \right] \right\} \right) \right].
\label{eq:FnPartnoT}
\end{equation}
Here, $H$ is the Hamiltonian (equation
(\ref{Eq:HSecondQ})) and $\overline{\eta},\eta$ are Grassmann variables
which are anti-commuting numbers \cite{grass}.
We recover the partition function itself with $Z [0,0]$.
The term $\sum_l \left[\overline{\eta}_l(\tau) a_l + a_l^{\dagger} \eta_l
(\tau) \right]$ is the generating functional which we will use to calculate
the correlation functions via functional derivatives. At the end of the
calculation, we will set $\overline{\eta} = \eta = 0$ to regain the correct
answer. Rewriting the above equation in functional integral form, we have

\begin{equation}
\fl Z\left[ \overline{\eta},\eta \right] = \int {\mathcal D}(\overline{\gamma},\gamma)
\exp \left\{-S+\int_0^{\beta} d\tau \sum_{l} \left[ \overline{\eta}_l(\tau)
\gamma_{l}(\tau) + \overline{\gamma}_{l}(\tau) \eta_l(\tau) \right] \right\},
\label{Eq:FnIZ}
\end{equation}
where the action is $S=\int_0^{\beta}
d\tau \sum_{l} \left[ \overline{\gamma}_{l}(\tau) \partial_{\tau} \gamma_{l}(\tau)
+ H(\overline{\gamma},\gamma) \right]$, and we have written $a_l$ and $a^\dagger_l$ in
terms of Grassmann variables $\gamma_{l}$ and $\overline{\gamma}_{l}$.
$H(\overline{\gamma},\gamma) = - \sum_{l=1}^{N-1} \left( J+j_l \right) \left(
\overline{\gamma}_l \gamma_{l+1} + \overline{\gamma}_{l+1} \gamma_l \right) -
B \sum_{l=1}^N \left( \textbf{1}-2 \overline{\gamma}_l \gamma_l \right)$ where
we have suppressed the $\tau$ argument after each term.
Next we take $n$ replicas of the partition functional, and average over the disorder $j_l$

\begin{equation}
\fl \overline{Z^n \left[ \overline{\eta},\eta \right]} = \frac{1}{(2\pi \Delta)^{N/2}}
\int_{-\infty}^\infty \left\{ \prod_{l=1}^N \left[ dj_l \exp \left( -\frac{1}{2\Delta} j_l^2 \right)
\right] \prod_{a=1}^n Z \left[ \overline{\eta}_a,\eta_a \right] \right\},
\end{equation}
where $Z \left[ \overline{\eta}_a,\eta_a \right]$ gives a subscript $a$ to every
term in equation (\ref{Eq:FnIZ}).
For each $l$, we can perform the average which corresponds to a  Hubbard-Stratonovich
transformation. This removes $j_l$ from
$\overline{Z^n \left[ \overline{\eta},\eta \right]}$ leaving $\Delta$. The
result of this is that we swap the non-translationally invariant $j_l$
term for a coupling between different replicas:

\begin{equation}
\fl \overline{Z^n \left[ \overline{\eta},\eta \right]} =  \int {\mathcal D}(\overline{\gamma},\gamma)
\exp \left\{-S^n +\int_0^{\beta} d\tau \sum_{l,a} \left[ \overline{\eta}_{l,a}(\tau)
\gamma_{l,a}(\tau) + \overline{\gamma}_{l,a}(\tau) \eta_{l,a}(\tau) \right] \right\},
\label{Eq:ZbarR}
\end{equation}
where $S = S_0^n - \Delta S_{int}^n$:

\begin{eqnarray}
\fl S_0^n & = & \int_0^{\beta} d\tau \sum_{l,a} \left[ \overline{\gamma}_{l,a} \partial_\tau
\gamma_{l,a} -J(\overline{\gamma}_{l,a}\gamma_{l+1,a}+\overline{\gamma}_{l+1,a}\gamma_{l,a}) +
2B \overline{\gamma}_{l,a} \gamma_{l,a} \right] \nonumber \\
\fl \Delta S_{int}^n & = & \frac{\Delta}{2} \sum_{l} \left\{ \int_0^{\beta} d\tau \sum_a \left[
\overline{\gamma}_{l,a} (\tau)\gamma_{l+1,a}(\tau)+\overline{\gamma}_{l+1,a}(\tau)
\gamma_{l,a}(\tau) \right] \right\}^2.
\end{eqnarray}
We can now diagonalise the unperturbed part of the Hamiltonian using a Fourier transform,
$\gamma_{l,a} (\tau) = \int_{-\pi}^\pi dq/(2\pi) \exp (i q l) \gamma_a (\tau,q)$.
Applying this transformation to equation \ref{Eq:ZbarR}, we find

\begin{eqnarray}
\fl S_0^n & = & \int_0^\beta d\tau \int_{-\pi}^\pi \frac{dq}{2\pi} \sum_a
\overline{\gamma}_a (\tau,q) \left[ \partial_\tau +\varepsilon (q) \right]\gamma_a (\tau,q) \nonumber \\
\fl \Delta S_{int}^n & = & \frac{\Delta}{2} \int_0^\beta d\tau d\tau' \int_{-\pi}^\pi
\prod_{m=1}^4 \left( \frac{dq_m}{2\pi} \right) \sum_{a,a'} 2\pi \delta (q_1-q_2+q_3-q_4)
\left( e^{iq_2} + e^{-iq_1} \right) \nonumber \\
\fl & \times & \left( e^{iq_4} + e^{-iq_3} \right)
\overline{\gamma}_a (\tau,q_1) \gamma_a (\tau,q_2) \overline{\gamma}_{a'} (\tau',q_3)
\gamma_{a'} (\tau',q_4),
\end{eqnarray}
and $\int_0^{\beta} d\tau \int_{-\pi}^\pi dq/(2\pi)
\sum_{a} \left[ \overline{\eta}_{a}(\tau,q)
\gamma_{a}(\tau,q) + \overline{\gamma}_{a}(\tau,q) \eta_{a}(\tau,q) \right]$,
remembering that $\varepsilon (q) = 2 (B-J\cos q)$.

In a system with no disorder and no
source terms, we could now calculate the partition function,
$ Z_0\left[0,0 \right]$. This calculation can
either be completed in the path integral formalism or directly from the
diagonalised Hamiltonian as found in equation \ref{Eq:PartFn0}.
We refer to, for example, \cite{CondMatFT} for the path integral calculation.

The next step is to define
$S'= \left[ \overline{\gamma} (\partial_{\tau} +\varepsilon)
\gamma - \overline{\eta}\gamma - \overline{\gamma}\eta \right]$,
where we have suppressed the equation's dependence on $\tau$, $q$ and $a$ for
simplicity.
Since $\mathcal{D} (\overline{\gamma}, \gamma )$ is invariant with respect to a
translation, we let $\gamma \rightarrow
\gamma+f$ and $\overline{\gamma} \rightarrow \overline{\gamma} + \overline{g}$
where $f$ and $g$ are fields that we can choose the value of
later. Substituting these new identities into $S'$ and then
expanding, we find that when $(\partial_{\tau} +\varepsilon)f=
\eta$ and $\overline{g} (-\overleftarrow{\partial}_{\tau} +\varepsilon)=\overline{\eta}$,
many terms disappear, and we are left with
$S' = \overline{\gamma} \left[ \partial_\tau +\varepsilon (q) \right] \gamma - \overline{\eta} f$.
Thus we need only find $f$ which we achieve by solving the inhomogeneous differential
equation above. Using Green functions,
$f(\tau) = \int_0^\beta d\tau' \mathcal{G} (\tau-\tau',q) \eta(\tau')$ where

\begin{equation}
\fl \mathcal{G} (\tau - \tau',q) = \exp \left[- \varepsilon (q) \left( \tau - \tau' \right) \right]
\left\{ \left[ 1-k(q) \right]\theta (\tau - \tau') - k(q) \theta (\tau' - \tau) \right\},
\label{Eq:Greensfn1}
\end{equation}
and $k(q)=(1+e^{\beta \varepsilon (q)})^{-1}$. When $\tau = \tau'$, we define the
Green function as $\mathcal{G} (0^-,q) =-k(q)$. Thus $S'$ becomes
$\int_0^\beta d\tau \int_{-\pi}^\pi \frac{dq}{2\pi} \sum_a \overline{\gamma}_a
(\tau,q) \left[ \partial_\tau +\varepsilon (q) \right] \gamma_a (\tau,q) -
\int_0^\beta d\tau d\tau' \int_{-\pi}^\pi \frac{dq}{2\pi} \sum_{a,b}
\overline{\eta}_a (\tau,q) \mathcal{G} (\tau-\tau',q) \eta_b (\tau',q)$.
The Green function term no longer depends on $\gamma$ and so can be taken
out of the integral:

\begin{equation}
\fl \overline{Z^n \left[ \overline{\eta}, \eta\right]} = \exp \left[\int_0^\beta d\tau d\tau'
\int_{-\pi}^\pi \frac{dq}{2\pi} \sum_{a,b} \overline{\eta}_a (\tau,q) \mathcal{G} (\tau-\tau',q)
\eta_b (\tau',q) \right] Z^n[0,0].
\end{equation}
Where $Z^n[0,0] = \int \mathcal{D} \left( \overline{\gamma}, \gamma \right) \exp \left(-S_0^n
+\Delta S_{int}^n \right)$.
We can also replace the $\gamma$s in the interaction term by taking advantage of
functional derivatives:

\begin{eqnarray}
\fl \Delta \widetilde{S}_{int}^n & = & \frac{\Delta}{2} \int_0^\beta d\tau d\tau' \int_{-\pi}^\pi
\prod_{m=1}^4 \left( \frac{dq_m}{2\pi} \right) \sum_{a,a'} 2\pi \delta (q_1-q_2+q_3-q_4)
\left( e^{iq_2} + e^{-iq_1} \right) \nonumber \\
\fl & \times &  \left( e^{iq_4} + e^{-iq_3} \right) 2\pi \frac{\delta}{\delta
\overline{\eta}_d (\tau',q_4)} 2\pi \frac{\delta}{\delta \eta_c
(\tau',q_3)} 2\pi \frac{\delta}{\delta \overline{\eta}_b (\tau,q_2)} 2\pi \frac{\delta}{\delta
\eta_a (\tau,q_1)}.
\end{eqnarray}
Perturbation theory requires an expansion in a small term. For us, this is small
term is $\Delta$. Since $\Delta = \overline{j_l^2}$, for the variance to be small,
the average of the square of the random part of the coupling strength must also be
small. Thus we expand $\exp (\Delta \widetilde{S}_{int}^n) =
( 1+\Delta \widetilde{S}_{int}^{n} +(\Delta \widetilde{S}_{int}^{n})^2/2+ \cdots )$.
We are left with

\begin{eqnarray}
\fl \overline{Z^n \left[ \overline{\eta}, \eta \right]} = \exp \left( -n\ln Z_0
\right) \exp \left( \Delta \widetilde{S}_{int}^n \right) \nonumber \\
\times \exp \left[ \int_0^\beta d\tau d\tau' \int_{-\pi}^\pi \frac{dq}{2\pi} \sum_{a,b} \overline{\eta}_a (\tau,q)
\mathcal{G}_{ab} (\tau-\tau',q) \eta_b (\tau',q) \right].
\end{eqnarray}
Thus we have found a useful form of the
disorder averaged replicated partition functional and are now in a position to
calculate the correlation functions we require.

The correlation functions are calculated by taking functional derivatives of the
partition functional, and then setting $\overline{\eta}=\eta=0$.

\begin{equation}
\fl \overline{\langle \mathcal{T} \psi (\tau_1, q_1) \overline{\psi} (\tau_2,q_2) \rangle} =
\left. \lim_{n \rightarrow 0} 2\pi \frac{\delta}{\delta \eta_1 (\tau_2,q_2)} 2\pi
\frac{\delta}{\delta \overline{\eta}_1 (\tau_1, q_1)}
\overline{Z^n[\overline{\eta},\eta]} \right|_{\overline{\eta}=\eta=0}
\label{Eq:QuZA}
\end{equation}
gives us a quenched disorder average, and

\begin{equation}
\fl \overline{\langle \mathcal{T} \psi (\tau_1, q_1) \overline{\psi} (\tau_2,q_2) \rangle} =
\left. \frac{1}{\overline{Z[\overline{\eta},\eta]}} 2\pi \frac{\delta}{\delta \eta_1 (\tau_2,q_2)} 2\pi
\frac{\delta}{\delta \overline{\eta}_1 (\tau_1, q_1)} \overline{Z[\overline{\eta},\eta]}
\right|_{\overline{\eta}=\eta=0}
\label{Eq:AnnZA}
\end{equation}
the annealed disorder average. Here, $\mathcal{T}$ is the time ordering operator.
Note that we need the average over the replicated
partition functional, $ \overline{Z^n [\overline{\eta},\eta]}$ for both,
and the limit $n \rightarrow 0$ in the quenched case means we are
effectively calculating the average of the log of the partition functional. This is
a variation of the replica method. We set $n=1$ for the annealed case.

In order to recover the needed correlation function, $\overline{\langle d^\dagger (q)
d(q) \rangle}$, and recalling that

\begin{equation}
\fl \overline{\langle \mathcal{T} \gamma (\tau,q) \overline{\gamma} (\tau',q') \rangle} =
\overline{\langle \gamma (\tau,q) \overline{\gamma} (\tau',q')
\rangle} \theta (\tau - \tau')-\overline{\langle \overline{\gamma} (\tau',q') \gamma
(\tau,q)  \rangle} \theta (\tau' - \tau),
\label{Eq:CorrFntau}
\end{equation}
we set $\tau = \tau'$ and $q=q'$. When the $\tau$s are equal, the time ordering gives
us $-\overline{\langle \overline{\gamma} (\tau,q) \gamma (\tau,q)  \rangle}$ which
can be rewritten as $- \overline{\langle d^\dagger (q) d(q) \rangle}$.

\begin{figure}[t]
\begin{center}
\centerline{
\includegraphics[width=1.9in]{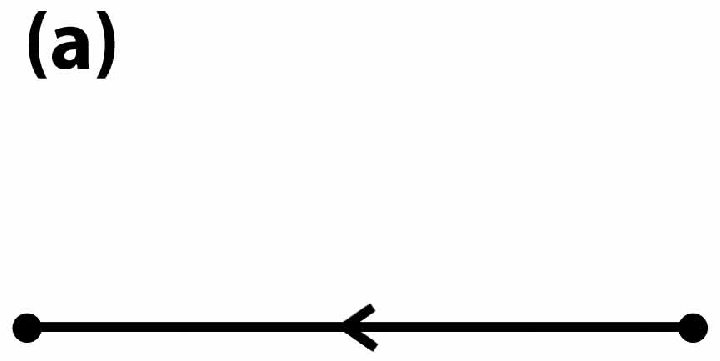}
\includegraphics[width=1.9in]{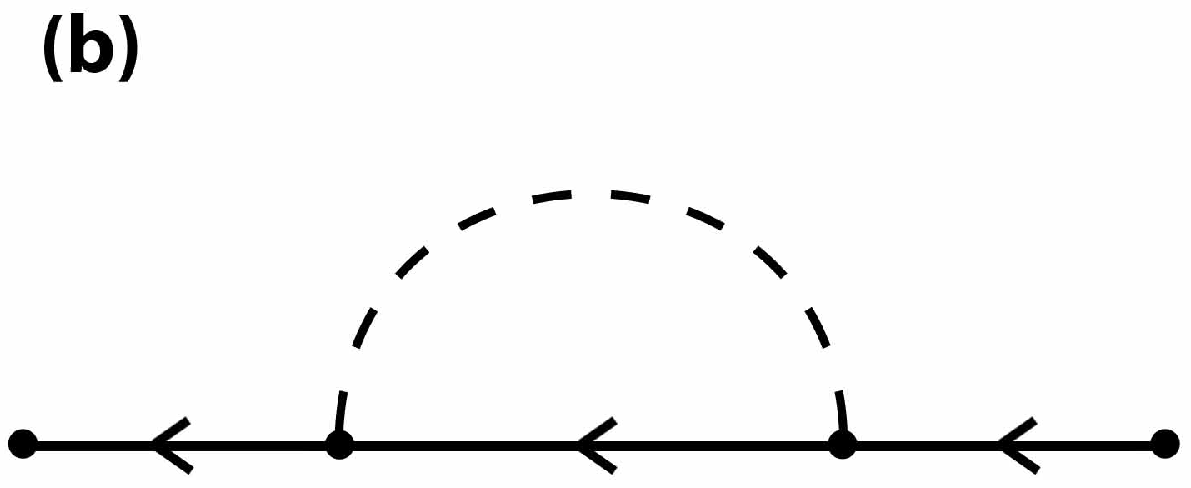}
\includegraphics[width=1.9in]{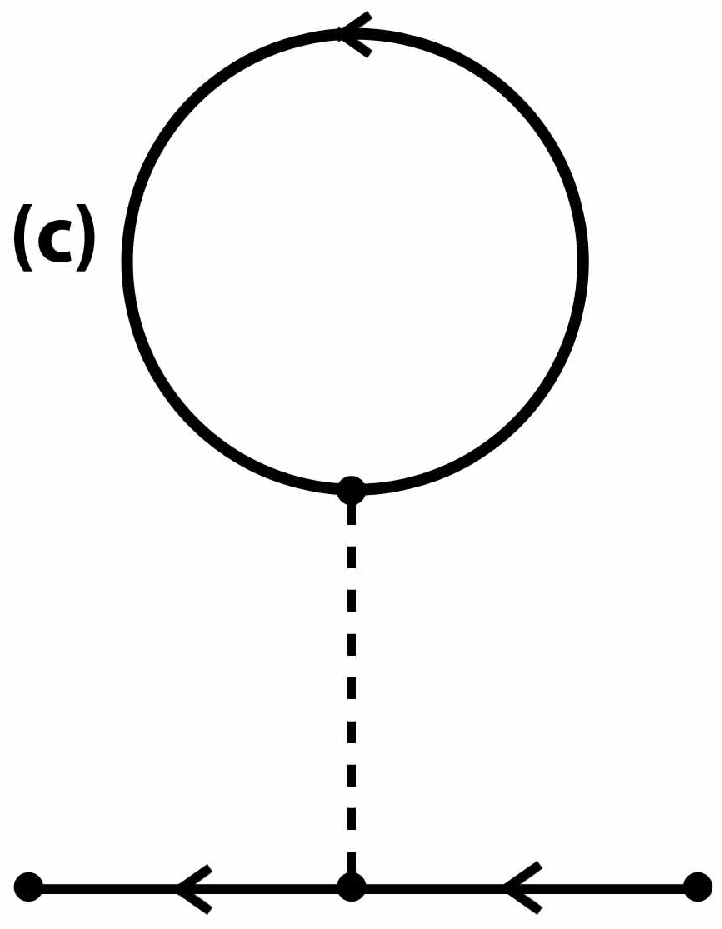}
 }
\end{center}
\caption{The calculated Feynman diagrams: (a) Is the unperturbed diagram, and (b) and
(c) are first order linked diagrams. Diagram (c) contributes only to the annealed average. }
\label{Fig:FeynDiagsabc}
\end{figure}

\section{Perturbation Theory And The Linked Cluster Theorem}

Feynman diagrams \cite{Mattuck} are an extremely useful tool which aid in
the calculation of complicated functional derivatives by converting them
into a diagrammatic form. These diagrams can be used to calculate the
perturbative expansion of the
correlation functions we require from $\overline{Z^n\left[\overline{\eta}, \eta \right]} $.
When constructing the possible Feynman diagrams of the disorder averaged correlation
functions, we will discover that both linked and vacuum
diagrams exist. Vacuum diagrams are those which contain two or more parts,
while a linked diagram is completely connected. Examples of
linked diagrams are shown in figures \ref{Fig:FeynDiagsabc}(b) and (c).

The linked cluster theorem demonstrates that when calculating correlation functions,
vacuum diagrams cancel out and only the linked diagrams contribute. The numerator
of the correlation functions is formed of a multiple of a sum of vacuum
diagrams and a sum of linked diagrams while the denominator is a sum of
vacuum diagrams only. The sums of vacuum diagrams therefore cancel, and we can
write the correlation function as a sum over the linked diagrams.

The linked cluster theorem also shows that the replica method
is exact in perturbation theory \cite{CondMatFT}. Taking the quenched limit is equivalent
to eliminating diagrams containing fermion loops (as shown in figure
\ref{Fig:FeynDiagsabc} (c)). Fermion loops arise from the average
introducing an interaction between different replicas. Prior to
the average, a diagram with a loop is actually a vacuum diagram.
In the quenched limit, these diagrams are no longer in the sum of
linked diagrams, instead appearing in the sum of vacuum diagrams.
Thus the quenched average leaves us a sum of linked diagrams without loops.

Since we will calculate both the quenched and annealed averages, we must find
and calculate all linked diagrams.

\section{Green Functions}

Equation \ref{Eq:GR1} shows that the correlation function
we need for the average is $\overline{\langle d^\dagger (q) d(q) \rangle}$ which is found from
equations \ref{Eq:QuZA} and \ref{Eq:AnnZA} for the quenched an annealed averages
respectively. To first order perturbation theory, the Feynman
diagrams contributing to the correlation function are shown in figure
\ref{Fig:FeynDiagsabc}. Using the standard rules for finite temperature
many-body perturbation theory, we can calculate the contribution of each diagram
(see, for example, \cite{Mattuck,Negele}).

For example, diagram figure \ref{Fig:FeynDiagsabc} (b) gives us

\begin{eqnarray}
\fl & & 2 \frac{\Delta}{2} \int_0^\beta d\tau d\tau' \int_{-\pi}^\pi \prod_{m=1}^4
\left( \frac{dq_m}{2\pi} \right) \sum_{abcd} \delta_{ab} \delta_{cd}
2\pi \delta \left( q_1-q_2+q_3-q_4 \right) (e^{iq_2}+e^{-iq_1}) \nonumber \\
\fl & \times & (e^{iq_4} + e^{-iq_3})2\pi \delta \left( q_\beta - q_3 \right) \mathcal{G} \left(
\tau_\beta - \tau' ,q_\beta \right) \delta_{\beta c}
2\pi \delta \left( q_4 - q_1 \right) \nonumber \\
\fl & \times & \mathcal{G} \left( \tau' - \tau , q_4 \right) \delta_{a d}
2\pi \delta \left( q_2 - q_\alpha \right) \mathcal{G} \left(
\tau - \tau_\alpha ,q_2 \right) \delta_{b \alpha},
\end{eqnarray}
where $\mathcal{G} \left( \tau - \tau' , q \right)$ was defined in equation
\ref{Eq:Greensfn1}. The
multiple of $2$ at the beginning indicates there are two topologically
distinct ways in which we can make this diagram. In order to recover the appropriate
correlation function, we set $q_\beta = q_\alpha$ and $\tau_\beta = \tau_\alpha$,
and perform the analytically possible integrals. This leaves us with the
$\cos ^2 [(p+q)/2]$ term in $\overline{G}_R$ below.

We find that the expansion of the correlation function gives

\begin{eqnarray}
\fl \overline{G}_R &  = &  \int_{-\pi}^\pi \frac{dq}{2\pi} \cos \left( q R
\right) \Bigg\{ \left[ 1 - 2k(q) \right] \\ \nonumber
\fl & + & 8 \Delta  \int_{-\pi}^\pi \frac{dp}{2\pi} \left[ \cos^2 \left( \frac{p+q}{2} \right)
\left\{ \frac{k(q)-k(p)}{\left[\varepsilon(p)-\varepsilon(q)\right]^2} - \frac{\beta k(q)
[1-k(q)]}{\varepsilon(p)-\varepsilon(q)}\right\} \right.  \\ \nonumber
\fl & - & \beta^2 n^2  \cos p \cos q k(p) \left[ 1-k(q) \right] k(q) \Bigg] \Bigg\}
\label{Eq:AveGR}
\end{eqnarray}
As discussed previously, for the quenched average, $n \rightarrow 0$ so the final
term in the above equation is zero while for the annealed average, $n=1$.

If we take the limit $\beta \rightarrow \infty$, we gain the zero temperature equivalent:

\begin{eqnarray}
\overline{G}_R & = & \frac{1}{R \pi} \left[ (1+e^{i\pi R}) \sin \left( \frac{\pi R}{2} \right)
-2 \sin \left( \alpha R \right) \right] \nonumber \\
 & + & 4 \Delta \int_0^{\pi/2} \frac{dp}{\pi} \int_0^\alpha
\frac{dq}{\pi} \left( \frac{\left[ 1+ \cos p \cos q\right] \left[ \cos (qR) - \cos (pR)
\right]}{\left[ \varepsilon (p) - \varepsilon (q) \right]^2} \right. \nonumber \\
 & + & \left. \frac{\left[ 1- \cos p
\cos q\right] \left[ \cos (qR) - e^{i\pi R}\cos (pR) \right]}{\left[ \nu (p) -
\varepsilon (q) \right]^2} \right)
\label{Eq:AveGRTZero}
\end{eqnarray}
where $\nu (x) = 2(B+J\cos x) $ and

$$
\alpha = \left\{ \begin{array}{lcl}
\cos^{-1} (B/J) & \mbox{for}& B < J \\
0 & \mbox{for} & B \geq J.
\end{array}\right.
$$
We note that the quenched and annealed averages are identical in the $T=0$ case.

We are now in a position to calculate the concurrence as described in previous
sections. In terms of $\overline{G}_R$, the nearest neighbour concurrence ($R=1$) is

\begin{equation}
\overline{\mathcal{C}}_1 = \left\{ 0, \left| \overline{G}_1 \right| - \frac{1}{2} \sqrt{
\left( 1+\overline{G}_0^2 -\overline{G}_1^2 \right)^2 - 4\overline{G}_0^2 } \right\},
\end{equation}
and the next nearest neighbour concurrence ($R=2$) is

\begin{equation}
\overline{\mathcal{C}}_2 = \left\{ 0, \left| \overline{G}_1^2 -\overline{G}_2 \overline{G}_0
 \right| - \frac{1}{2} \sqrt{\left( 1+\overline{G}_0^2 -\overline{G}_2^2 \right)^2 -
 4\overline{G}_0^2 } \right\}.
\end{equation}

Then using equations 7.1 (and \ref{Eq:AveGRTZero} for zero temperature), we can find
the quenched and annealed averages of the concurrence.

We also plan to briefly discuss the single-site entanglement entropy,
$\overline{S}(\rho_l) = -\{(1+\overline{G}_0) \log_2 [(1+
\overline{G}_0)/2] +(1-\overline{G}_0) \log_2 [(1-\overline{G}_0)/2]\}/2$ which measures
entanglement at zero temperature between a single site and the rest of the spin chain.
In addition we discuss the two-site entanglement entropy, $\overline{S}(\rho_{l,l+1}) =
-a \log a - c \log c - d \log [(b+d)/(b-d)] - b \log (b^2-d^2)$ where $a = [(1+
\overline{G}_0)^2-\overline{G}_1^2]/4$, $b = (1-\overline{G}_0^2 +\overline{G}_1^2)/4$,
$c = [(1-\overline{G}_0)^2-\overline{G}_1^2]/4$ and $d = -\overline{G}_1/2$, which again
can be used at zero temperature, and measures the entanglement between the two sites and the
rest of the chain.

\begin{figure}[t]
\begin{center}
\centerline{
\includegraphics[width=4in]{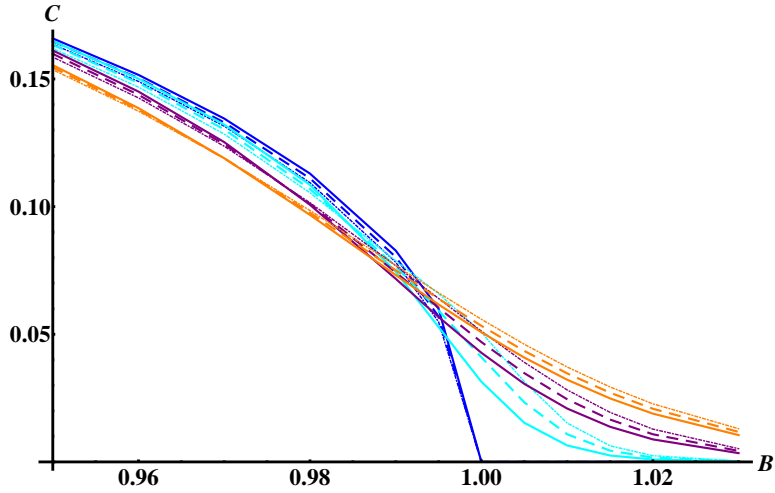}
 }
\end{center}
\caption{Quenched nearest neighbour concurrence around the quantum critical point.
The solid lines are no
disorder, the longer dashes $\Delta=0.0005$ and the short dashes $\Delta= 0.001$.
The blue lines are $T=0$, the cyan lines $T=0.01$, the purple lines $T=0.02$
and the orange lines $T=0.03$.}
\label{fig:QuenchedConcQPTR1}
\end{figure}

\section{Disorder Averaged Concurrence}

We have restricted the plots to be close to the quantum critical point (QCP) which
for this system occurs at $B=J$ since this is the region where the effect of
the disorder is the most interesting.

Figure \ref{fig:QuenchedConcQPTR1} demonstrates how the quenched nearest neighbour
concurrence behaves as the magnetic field varies for different values of
temperature and disorder.
We find that at zero temperature, increasing $\Delta$ always decreases the concurrence,
and no new entanglement is created above the QCP. For finite temperatures,
there is a crossover point of the
magnetic field, $B_c$, below which increasing $\Delta$ decreases concurrence, and
above which, increasing $\Delta$ increases concurrence. As the temperature increases,
$B_c$ generally decreases while the amount that $\Delta$ increases the concurrence
above $B_c$ decreases.

\begin{figure}[t]
\begin{center}
\centerline{
\includegraphics[width=4in]{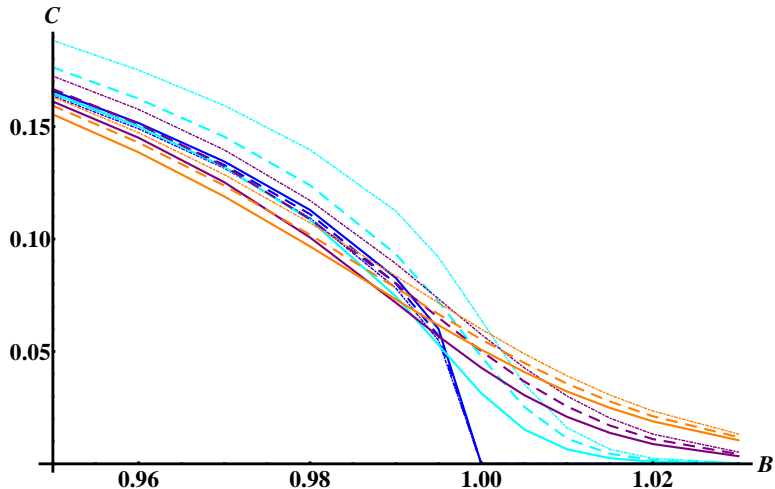}
 }
\end{center}
\caption{Annealed nearest neighbour concurrence around the quantum critical point.
The solid lines are no
disorder, the longer dashes $\Delta=0.0005$ and the short dashes $\Delta= 0.001$.
The blue lines are $T=0$, the cyan lines $T=0.01$, the purple lines $T=0.02$
and the orange lines $T=0.03$.}
\label{fig:AnnealedConcQPTR1}
\end{figure}

\begin{figure}[t]
\begin{center}
\centerline{
\includegraphics[width=4in]{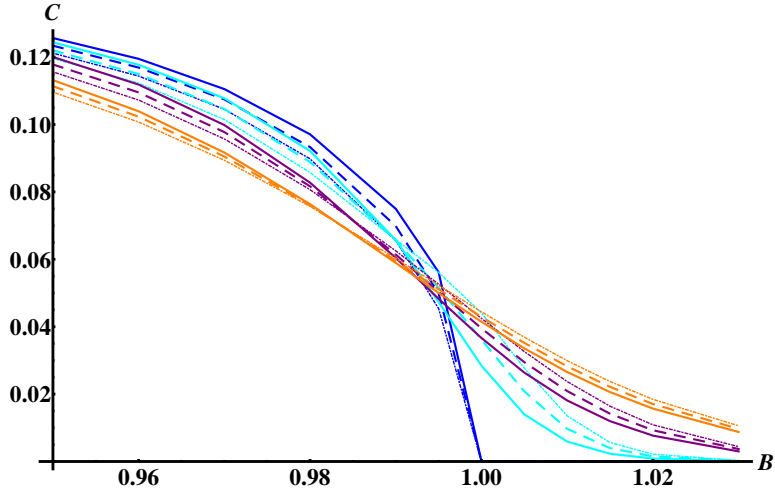}
 }
\end{center}
\caption{Quenched next nearest neighbour concurrence around the quantum critical point.
The solid lines are no
disorder, the longer dashes $\Delta=0.0005$ and the short dashes $\Delta= 0.001$.
The blue lines are $T=0$, the cyan lines $T=0.01$, the purple lines $T=0.02$
and the orange lines $T=0.03$.}
\label{fig:QuenchedConcQPTR2}
\end{figure}

\begin{figure}[t]
\begin{center}
\centerline{
\includegraphics[width=4in]{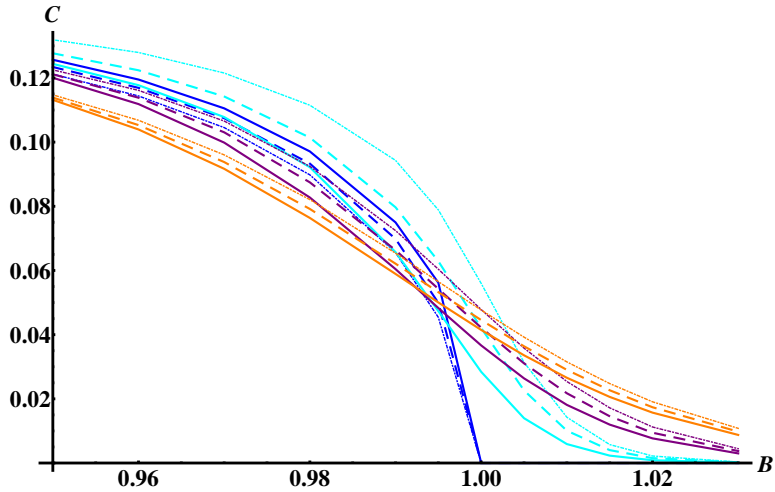}
 }
\end{center}
\caption{Annealed next nearest neighbour concurrence around the quantum critical point.
The solid lines are no
disorder, the longer dashes $\Delta=0.0005$ and the short dashes $\Delta= 0.001$.
The blue lines are $T=0$, the cyan lines $T=0.01$, the purple lines $T=0.02$
and the orange lines $T=0.03$.}
\label{fig:AnnealedConcQPTR2}
\end{figure}

For the annealed nearest neighbour concurrence, shown in figure \ref{fig:AnnealedConcQPTR1},
we find a similar pattern of behaviour. However, $B_c$ in this case is much lower than
for the quenched average.
Both the quenched and annealed next nearest neighbour concurrence, shown in figures
\ref{fig:QuenchedConcQPTR2} and \ref{fig:AnnealedConcQPTR2} respectively again give similar
results, but with lower values for the concurrence overall.

A possible reason for the behaviour described above could be that the disorder has a
similar effect on entanglement to the temperature. Both have the effect
of mixing energy levels allowing for the possibility of creating entanglement above
the QCP. However, at higher temperatures and lower magnetic field, the likely
effect of more mixing is to decrease entanglement.

A puzzling feature of these results is that disorder does not increase the concurrence
above the QCP at zero temperature as we would expect following the reasoning in the
argument above. We expect that we would need to calculate the perturbation
series to a higher order to see this behaviour. Alternatively, perhaps the value
of $\Delta$ simply needs to be larger than perturbation theory is valid for to
show this.

We have also calculated the correlation functions for a random magnetic field, when
$H_1(b_l) = \sum_l b_l \sigma_l^z$. Again we find the same behaviour for the concurrence,
with a crossover point, $B_c$, close to the QCP for the quenched average and
lower for the annealed average, and zero concurrence at $T=0$ for $B \geq J$
for both nearest and next nearest neighbour concurrence. Close to the QCP,
increasing $\Delta$ for $b_l$, increases the concurrence more than for $j_l$ for
both nearest and next nearest neighbour entanglement.

In addition to the concurrence we have considered the single and two site entanglement
entropy. Since $\overline{G}_0 = G_0$ at zero temperature, disorder doesn't affect
the single-site entanglement entropy in the regime we are able to apply perturbation
theory in. This is the case for whether the randomness is in $j_l$ or $b_l$.

Interestingly, the two-site
entanglement entropy actually increases with disorder for both $j_l$ and $b_l$
for all values of $B$. Increasing $\Delta$ increases the amount of entanglement
by a larger amount the closer $B$ gets to $J$ for each. The random magnetic field,
$b_l$, has a greater effect on the entanglement that $j_l$; increasing $\Delta$
by the same amount increases the entanglement entropy more for $b_l$ than for $j_l$.

We compare our results to \cite{random1} which uses sampling to look at
an $XX$ spin chain in a random magnetic field. They consider concurrence
at zero temperature, and the random field taken from a Gaussian distribution
as well as a Lorentzian distribution. They find that increasing the disorder
below the QCP decreases the nearest neighbour concurrence, while above
the QCP, increasing disorder increases the concurrence. While our results
for zero temperature agree with this behaviour below the QCP, above
$B=J$, concurrence remains zero for us. However, we consider weak
disorder while the lowest disorder \cite{random1} calculates is an
order of magnitude larger than ours. It is possible that calculating the perturbative
expansion to higher order would allow us to observe this behaviour.

\section{Conclusions}

We have found that for weak disorder, concurrence, the entanglement between
qubit pairs, in general decreases with increasing quenched disorder. At non-zero temperatures,
close to the QCP, disorder instead acts to increase entanglement. In the case
of two-site entanglement entropy, (remembering this is a measure of entanglement
only at zero temperature), the entanglement between two sites and the
remainder of the spin chain, increasing the disorder increases entanglement.

Our results demonstrate that disorder is not necessarily detrimental
to entanglement and thus to schemes which use it as a resource. This
is, however, dependent on which entanglement
measure is appropriate and the parameter regime we consider.

We can use the perturbation theory method to find the disorder averaged
concurrence for other systems. However, if the system cannot be expressed as
a free fermion (non-interacting) model, the number of disorder averaged
correlation functions required would be increased since Wick's theorem
could no longer be applied. For example, we would no longer have
$ \overline{\langle \sigma_l^z \sigma_{l+1}^z \rangle} = \overline{G}_0^2 -
\overline{G}_R^2$, and would instead need to calculate a four-point
correlation function since $ \overline{\langle \sigma_l^z \sigma_{l+1}^z \rangle} =
2\overline{G}_0 - 1 + 4 \overline{\langle a_l^\dagger a_l a_{l+R}^\dagger a_{l+R}
\rangle}$.

One strength of our paper is that we consider finite temperature as
well as zero temperature entanglement.
Since we use perturbative methods, our work has the propensity to be
extended in many directions. For example, higher order calculations and
application of the Dyson equation would be useful. In particular,
perturbation theory allows for the consideration of
time dependent non-equilibrium disorder averaged measures of entanglement
using techniques such as the Keldysh formalism. This
would be interesting since it would show how disorder affects the finite
temperature entanglement of a system as it evolves over time.

\end{document}